\documentclass[%draft,
superscriptaddress,reprint,amsmath,amssymb,aip,pra,rsi, floatfix]{revtex4-1}
\usepackage[margin=1.8cm]{geometry}
\usepackage{xcolor} %for colored text in comments
\usepackage{graphicx}% Include figure files
\usepackage{dcolumn}% Align table columns on decimal point
\usepackage{upgreek}% Allows text micron
\usepackage{bm}% bold math
\usepackage{hyperref}% add hypertext capabilities
\usepackage{times}
\usepackage[mathlines]{lineno}% Enable numbering of text and display math
\usepackage{physics}
\usepackage{cancel} % pretty strikethrough
\usepackage{soul}
\usepackage{epstopdf}

\usepackage{etoolbox}
\makeatletter
\newcommand*{\newbibstartnumber}[1]{%
  \apptocmd{\thebibliography}{%
    \global\c@NAT@ctr #1\relax
    \addtocounter{NAT@ctr}{-1}%
  }{}{}%
}
\makeatother

\usepackage{graphicx}% Include figure files
\usepackage{float}
\usepackage{dcolumn}% Align table columns on decimal point
\usepackage{bm}% bold math
\usepackage{breqn}
\usepackage{braket}
\usepackage{mathtools}

\makeatletter
\let\cat@comma@active\@empty
\makeatother

\makeatletter
\let\@fnsymbol\@fnsymbol@latex
\@booleanfalse\altaffilletter@sw
\makeatother

\begin{document}

\title{Entanglement Across Separate Silicon Dies in a Modular Superconducting Qubit Device}

\author{Alysson Gold}
%\thanks{Electronic mail: agold@rigetti.com}
\thanks{These two authors contributed equally. Corresponding electronic mail: agold@rigetti.com}
\affiliation{Rigetti Computing, 775 Heinz Ave, Berkeley CA 94701}
\author{JP Paquette}
\thanks{These two authors contributed equally. Corresponding electronic mail: agold@rigetti.com}
\affiliation{Rigetti Computing, 775 Heinz Ave, Berkeley CA 94701}
\author{Anna Stockklauser}
\affiliation{Rigetti Computing, 775 Heinz Ave, Berkeley CA 94701}
\author{Matthew J. Reagor}
\affiliation{Rigetti Computing, 775 Heinz Ave, Berkeley CA 94701}
\author{M. Sohaib Alam}
\affiliation{Rigetti Computing, 775 Heinz Ave, Berkeley CA 94701}
\author{Andrew Bestwick}
\affiliation{Rigetti Computing, 775 Heinz Ave, Berkeley CA 94701}
\author{Nicolas Didier}
\affiliation{Rigetti Computing, 775 Heinz Ave, Berkeley CA 94701}
\author{Ani Nersisyan}
\affiliation{Rigetti Computing, 775 Heinz Ave, Berkeley CA 94701}
\author{Feyza Oruc}
\affiliation{Rigetti Computing, 775 Heinz Ave, Berkeley CA 94701}
\author{Armin Razavi}
\affiliation{Rigetti Computing, 775 Heinz Ave, Berkeley CA 94701}
\author{Ben Scharmann}
\affiliation{Rigetti Computing, 775 Heinz Ave, Berkeley CA 94701}
\author{Eyob A. Sete}
\affiliation{Rigetti Computing, 775 Heinz Ave, Berkeley CA 94701}
\author{Biswajit Sur}
\affiliation{Rigetti Computing, 775 Heinz Ave, Berkeley CA 94701}
\author{Davide Venturelli}
\affiliation{Quantum Artificial Intelligence Laboratory (QuAIL), NASA Ames Research Center, Moffett Field, CA 94035, USA}
\affiliation{USRA Research Institute for Advanced Computer Science (RIACS), Mountain View, CA 94043, USA}
\author{Cody James Winkleblack}
\affiliation{Rigetti Computing, 775 Heinz Ave, Berkeley CA 94701}
\author{Filip Wudarski}
\affiliation{Quantum Artificial Intelligence Laboratory (QuAIL), NASA Ames Research Center, Moffett Field, CA 94035, USA}
\affiliation{USRA Research Institute for Advanced Computer Science (RIACS), Mountain View, CA 94043, USA}
\author{Mike Harburn}
\affiliation{Rigetti Computing, 775 Heinz Ave, Berkeley CA 94701}
\author{Chad Rigetti}
\affiliation{Rigetti Computing, 775 Heinz Ave, Berkeley CA 94701}

\date{\today}% It is always \today, today,

\begin{abstract}
Assembling future large-scale quantum computers out of smaller, specialized modules promises to simplify a number of formidable science and engineering challenges. One of the primary challenges in developing a modular architecture is in engineering high fidelity, low-latency quantum interconnects between modules. Here we demonstrate a modular solid state architecture with deterministic inter-module coupling between four physically separate, interchangeable superconducting qubit integrated circuits, achieving two-qubit gate fidelities as high as 99.1$\pm0.5$\% and 98.3$\pm$0.3\% for iSWAP and CZ entangling gates, respectively. The quality of the inter-module entanglement is further confirmed by a demonstration of Bell-inequality violation for disjoint pairs of entangled qubits across the four separate silicon dies. Having proven out the fundamental building blocks, this work provides the technological foundations for a modular quantum processor: technology which will accelerate near-term experimental efforts and open up new paths to the fault-tolerant era for solid state qubit architectures.
\end{abstract}

\maketitle

\section{\label{sec:introduction} Introduction}

Progress in quantum operations over multi-node networks could enable modular architectures spanning distances from the nanometer to the kilometer scale \cite{PRXQuantum.2.017002, steffen2013deterministic, chou2018deterministic, wan2019quantum, hensen2015loophole}. Heralded entanglement protocols, whereby entanglement is generated probabilistically, have now reached entanglement rates up to 200 Hz \cite{olmschenk2009quantum, moehring2007entanglement, humphreys2018deterministic, monroe2014large, ritter2012elementary, Zhong_2020,krastanov2021}. Superconducting systems have established direct exchange of quantum information over cryogenic microwave channels \cite{PhysRevLett.120.200501, PhysRevLett.125.260502, axline2018demand, leung2019deterministic, zhong2020deterministic, kurpiers2018deterministic}, which is particularly useful towards interconnects of intermediate range such as between dilution refrigerators. Yet, in the context of superconducting qubit based processors, none of these methods are likely to outperform local gates between qubits, which can achieve coupling rates in the tens of MHz and fidelities reaching 99.9\% \cite{Chen_2020, Foxen_2020, negirneac2020highfidelity, sung2020realization, stehlik2021tunable}. Importantly, modules consisting of closely spaced and directly coupled separate physical dies retain many of the benefits of distributed modular architectures without the challenge of remote entanglement. Increased isolation between modules reduces cross-talk and correlated errors, for example due to high energy background radiation \cite{wilen2020correlated, vepsalainen2020impact, cardani2020reducing}, and by fabricating smaller units and selecting the highest yielding units for device assembly, higher device yield is achievable \cite{8614500, dickel2018chip}. Mastering 3D integration and modular solid state architectures has thus been a long-standing objective \cite{rosenberg20173d, foxen2017qubit, brecht2016multilayer}.

We demonstrate herein a modular superconducting qubit device with direct coupling between physical modules. The device, which consists of four eight-qubit integrated circuits fabricated on individual dies and flip-chip bonded to a larger carrier chip, achieves coupling rates and entanglement quality similar to the state-of-the-art in intra-chip coupling.

\section{\label{sec:Results} Results}
\subsection{\label{sec:des} Design of a Modular Superconducting Qubit Device}

\begin{figure*}
    \centering
    \includegraphics[width=\textwidth]{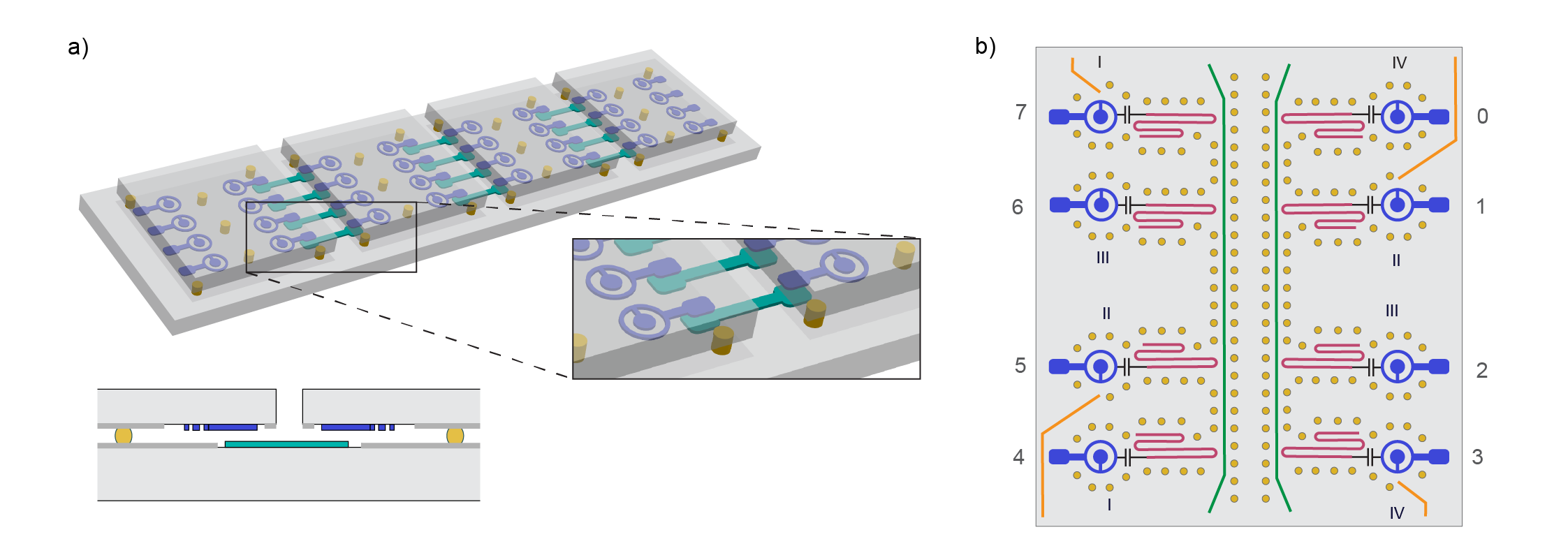}
    \caption{
    \label{fig:multiDieSchem}
    (a) Isometric %and (b) Cross-sectional 
    view of the device assembly. The qubits (blue circular structures) are fabricated on the QuIC die and have one arm with a paddle-shaped coupler extending to the edge of the chip. The chips are flip-chip bonded onto the carrier chip using indium bump bonds (yellow) and the qubit couplers are aligned above couplers on the carrier chip (teal) as shown in the inset as well as the cross-sectional view. (b) False-colored image of a single QuIC including readout resonators and readout lines (magenta and green), indium bumps (yellow), flux bias lines (orange) and the qubits and paddles of the inter-chip couplers (blue). The physical qubits are labelled 0-7 while Roman numerals correspond to the design specification for the qubit (see Methods).}
\end{figure*}

The multi-die device assembly is constructed through flip-chip bonding of four nominally identical dies to a larger carrier die as shown in Fig.~\ref{fig:multiDieSchem}a. The carrier chip assumes a similar role to the chip multiprocessor in a classical multi-core processor while also providing microwave shielding, circuitry to interface between the individual QuIC chips and signal routing for the device I/O. The smaller individual dies comprise the qubit integrated circuits (QuIC), each consisting of four flux tunable and four fixed transmon qubits \cite{reagor2018demonstration}, and corresponding readout resonators and flux control lines as shown in Fig.~\ref{fig:multiDieSchem}b. The readout is multiplexed with four qubits and resonators per readout line and qubits are driven through the readout on this test platform. Qubits are labelled with a letter for the die position from left to right and a number for the qubit position within the die, e.g. B6. Entangled pairs are labeled according to the adjacent qubits, eg. B6-C1. The QuIC dies are designed to be identical in order to maximize fabrication yield and enable modular assembly. The benefits to fabrication yield are evident in considering the number of distinct permutations that exist for a single device assembly: for a wafer with 220 QuIC dies, there are over $2.2\mathrm{E}9$ possible unique device assemblies.

The device Hamiltonian is designed to enable two-qubit parametric gates \cite{bertet2006parametric, niskanen2007quantum, didier2018parametric, reagor2018demonstration, hong2020demonstration} between pairs of qubits on separate dies (see Methods). Coupling between qubits on separate chips is mediated through capacitive couplers on the QuIC and the carrier side, resulting in a cross-chip, charge-charge interaction. The carrier chip contains couplers with paddles at each end which are positioned below similar paddle-shaped couplers extending from the qubits on the QuIC as shown in Fig.~\ref{fig:multiDieSchem}a. There is no coupling between qubits on the same die in this test platform to isolate the basic inter-chip coupling mechanism and avoid complexities arising from larger circuits such as frequency collisions and leakage. However, the qubit and coupler design can be adapted to a larger lattice with intra-chip connectivity. 

\subsection{\label{sec:fab} Device Fabrication, Assembly, and Validation}
The QuIC chips are fabricated using standard lithographic techniques on a Si wafer which is then diced to create individual dies. The Josephson junctions which form the SQUID loops of the transmon qubits are fabricated through double-angle evaporation of Al. Superconducting circuit components, including the Al pads for the Josephson junctions and Nb ground planes, signal routing and coplanar waveguides and resonators, are fabricated by pattern, deposition and liftoff steps \cite{nersisyan2019manufacturing}.

Flip-chip bonding of the carrier and QuIC modules is accomplished through the deposition and patterning of indium bumps of height $6.5\ \mu \mathrm{m}$ and $40\ \mu\mathrm{m}$ diameter onto the carrier chip. The QuIC chips are flipped and aligned to the carrier before thermo-compression bonding, creating a superconducting bond between the carrier and QuIC chips. The fabrication process is described in detail in the Methods section. The indium bump heights post-bonding determine the vertical separation between the QuIC chips and carrier, as shown in Fig.~\ref{fig:multiDieSchem}a.

\begin{figure}
    \centering
    \includegraphics[width=\columnwidth]{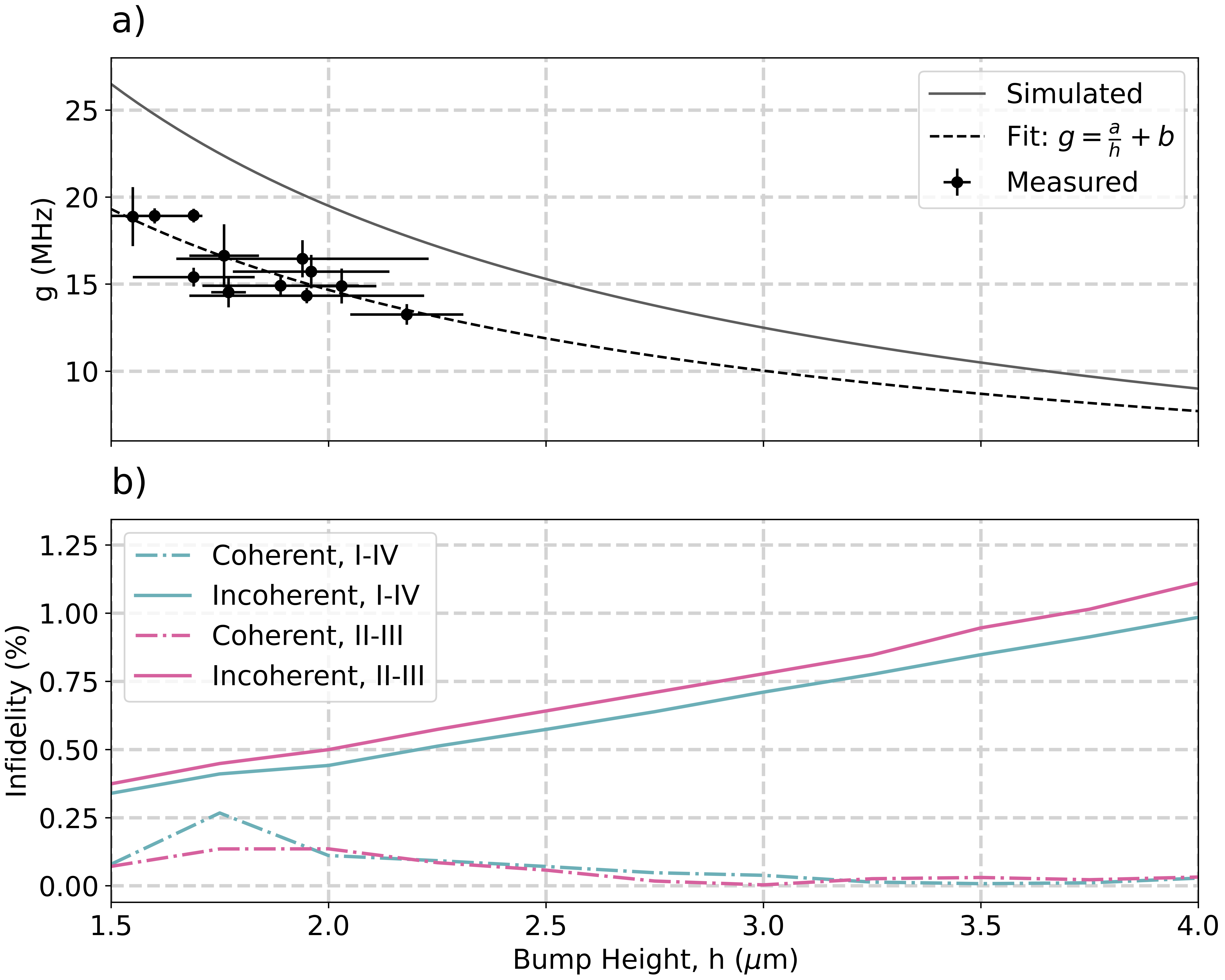}
    \caption{Impact of post-bonding indium bump height on a) coupling rate, $g$ in linear units and b) both coherent and incoherent simulated gate errors. The fit parameters in a) are $a=27.9\pm6.4$ MHz $\mu$m and $b=0.7\pm3.5$ MHz. The discrepancy between measurement and simulation and details on bump height measurements are discussed in the Methods section. There are only two distinct entangled qubit pairs in the designed Hamiltonian, the II-III and I-IV pairs as shown in Fig.~\ref{fig:multiDieSchem}b, both of which are simulated in b).}
    \label{fig:couplingSummary}
\end{figure}

Importantly, the capacitance between the carrier and QuIC paddles is inversely proportional to the height of the indium bump bonds, $h$, as expected for a parallel plate capacitor. The bare coupling rate between qubits, $g$ is directly proportional to this capacitance and thus follows the same dependence on $h$. Due to bonding process variation, the indium bump height spans a range of 1.5 $\mu$m to 4 $\mu$m. As shown in Fig.~\ref{fig:couplingSummary}, this corresponds to a range for $g/2\pi$ of 8.8 MHz to 26.1 MHz for the coupling rate.
 
Despite the range of anticipated couplings, the simulated fidelity for parametric gates in this design is relatively unaffected. Figure~\ref{fig:couplingSummary} shows simulation results of both the parametric CZ unitary gate error in the absence of loss and dephasing channels (coherent error) and the coherence-limited gate error (incoherent error) as a function of the bump height. The incoherent error is obtained assuming an ideal coherent exchange between the qubits while the coherent error takes into account the unwanted interactions arising from the capacitive coupling and flux modulation. For the coherence-limited fidelity calculation, we use a relaxation time, $\mathrm{T}_1$, of 73/18 $\mu$s and a dephasing time, $\mathrm{T}_2$, of 43/15 $\mu$s for fixed/tunable qubits. Over the full range of indium bump height expected from the bonding process, the predicted maximum achievable fidelity (taken as the minimum of the coherence-limited and unitary fidelity) varies from just under 99.0\% to 99.5\%. For an initial proof of concept, this range is acceptable; however, to push towards fidelities exceeding 99\% or to employ this as part of a tunable coupler scheme \cite{PhysRevLett.125.240503, PhysRevApplied.10.054062}, efforts will be needed to reduce the spread. Additional calibration of the force applied during the bonding process and design revisions to reduce the sensitivity of the coupling to bump height by changing the paddle geometry could reduce this variation further for gate schemes requiring a tighter tolerance.

The device assembly, designed and fabricated as described above, is measured in a dilution refrigerator at 10 mK. To assess the accuracy of the simulations and modeling conducted during device design, we characterize the device Hamiltonian (qubit and resonator frequencies, and coupling rates between elements) and compare with predictions from simulation. Qubit frequencies are within 2.1\% of predicted values for the $f_{01}$ transition and 11\% for the qubit anharmonicity at zero applied flux bias (see Methods), demonstrating good agreement and indicating the inter-chip coupling technology does not impact the steady-state device physics in an unexpected manner.

\subsection{\label{app:g_meas} Experimental Determination of the Bare Coupling Rates}
The capacitive coupling formed by the inter-chip couplers results in a charge-charge interaction between the coupled qubits, $q_1$ and $q_2$. In the dispersive regime, where the detuning between qubits is large compared to the bare coupling rate, $|\omega_{01, 1}-\omega_{01, 2}| \gg g$, we can calculate the bare coupling rate from the measured dispersive shift, $\chi_{qq}$. The relationship between $\chi_{qq}$ and $g$ is given by Eq.~\eqref{eq:qqchi} for the general case of two flux-tunable transmons, which differs from the treatment of transmon-resonator dispersive shifts. Note here we are working in the transmon limit and the equation below is the result of a perturbative expansion of the Hamiltonian eigenstates and eigenenergies as a function of applied magnetic flux, following the treatment in Ref.~\onlinecite{didier2018parametric}. $\mathrm{E}_{\mathrm{J, eff}}(\Phi)$ is the effective Josephson energy of the DC SQUID, a function of the applied magnetic flux through the SQUID, $\Phi$, and is defined, along with $\lambda(\Phi)$ and $\Lambda(\Phi)$ in the same reference.
\begin{align}
\label{eq:qqchi}
\chi_{qq}(\Phi_1, \Phi_2) &= 2g^2 \bigg[\frac{\mu_{01,1}^2(\Phi_1) \mu_{12,2}^2(\Phi_2) }{\omega_{01,1}(\Phi_1) - \omega_{12,2}(\Phi_2)}\nonumber\\ &\qquad-\frac{\mu_{12,1}^2(\Phi_1) \mu_{01,2}^2(\Phi_2)}{\omega_{12,1}(\Phi_1) - \omega_{01,2}(\Phi_2)}\bigg],\\
    \mu_{01}(\Phi) &= \left[\frac{\mathrm{E}_{\mathrm{J, eff}}(\Phi)}{\mathrm{E}_{\mathrm{J, eff}}(0)}\right]^{1/4} \frac{\lambda(\Phi)}{\lambda(0)},\\
    \mu_{12}(\Phi) &= \left[\frac{\mathrm{E}_{\mathrm{J, eff}}(\Phi)}{\mathrm{E}_{\mathrm{J, eff}}(0)}\right]^{1/4} \frac{\Lambda(\Phi)}{\lambda(0)}.
\end{align}

We measure the dispersive shift through time Ramsey measurements. In a time Ramsey measurement, an X/2 pulse is applied to a qubit, $q_1$, causing the qubit to precess about the equator. After a time delay, $\Delta t$, a Z pulse rotates the qubit through a phase $\phi = 2\pi\Delta t\delta f$, where $\delta f$ is the detuning of the pulse frequency relative to the qubit frequency, $f_{01}$. Finally, another X/2 pulse is applied and the qubit state is measured. The resulting excited state visibility oscillates as a function of the time delay, reaching full visibility when the Z rotation perfectly offsets the phase accumulated from the precession during the delay time. From the period of the oscillations, the difference between the qubit frequency and the applied pulse frequency (already detuned from the expected qubit frequency by $\delta f$), and by consequence the qubit frequency itself as the applied pulse frequency is well defined by the control electronics, can be determined with high precision. 

To measure $\chi_{qq}$, the time Ramsey measurement is performed on a qubit $q_1$ with adjacent qubit $q_2$ in the ground state and again in the excited state. The difference between the $f_{01}$ measured for $q_1$ at both points gives a precise measurement of $\chi_{qq}$. The measurement is then performed in the opposite direction, with the state of $q_1$ varied while the frequency of $q_2$ is measured. The bare coupling rate calculated from the dispersive shift, as given by Eq.~\ref{eq:qqchi}, should be equivalent in both directions for a pure ZZ interaction, to within the error of the measurement. While the design target for all couplers was 12 MHz for a 3 $\mu$m indium bump height post-bonding, the observed coupling rate varied across the chip from 13.26$\pm$0.59 MHz to 18.94$\pm$0.39 MHz (see Fig.~\ref{fig:couplingSummary}). This was within the anticipated range due to indium bump height variation (see Methods for further details, in particular Table~\ref{tab:CouplingStrengths}).

\subsection{\label{sec:DevChar} Cross-chip Entangling Gate Performance}

We calibrated and benchmarked gates on ten out of twelve inter-chip pairs. The remaining two pairs could achieve population transfer but due to frequency targeting error in the fabrication process, the gate modulation frequencies were outside of the frequency band of the control electronics and degraded AC flux control resulted in low fidelity. The primary benchmarking methods employed were two-qubit randomized benchmarking (RB) and interleaved randomized benchmarking (iRB) \cite{Magesan_2012, knill2008randomized}. We quote the estimate from iRB when the RB protocol estimates an average gate fidelity of $\mathcal{F}_{\mathrm{RB}}\geq 92\%$, which bounds the iRB estimate to at most 20\% of the reported gate error due to imperfect gate randomization \cite{Magesan_2012}. Below this empirical threshold, the assumptions of the error model can lead to large uncertainty and an overestimate of the gate fidelity for iRB. 

Table~\ref{tab:Coherence Limited Fidelity} provides a summary of the CZ gate fidelities measured for each of the ten pairs and compares them with the coherence limited fidelity (the maximum achievable fidelity predicted from the measured relaxation and dephasing times of the qubit pair). The coherence limited fidelity is computed from the $\mathrm{T}_1$ and $\mathrm{T}_2$ under modulation, $\widetilde{\mathrm{T}}_1$ and $\widetilde{\mathrm{T}}_2$, ie. the coherence times as measured while an AC flux bias is applied to the tunable qubit at the gate modulation frequency, emulating conditions during gate operation. $\widetilde{\mathrm{T}}_1$ and $\widetilde{\mathrm{T}}_2$ for the tunable qubit in each pair, the limiting qubit in regards to coherence, are also recorded in the table.

Comparing the measured fidelities with the coherence limited fidelities, the fidelity is almost always limited by the qubit coherence suggesting the inter-chip coupling mechanism does not limit gate error directly. Furthermore, we have compared qubit coherence times for inter-chip-coupled qubits to a baseline of similar qubits that are not coupled by inter-chip couplers and coherence times do not appear to be limited by the inter-chip coupling technology itself (see Methods). This suggests that gate fidelities are limited by the same sources of error arising in monolithic devices and are not directly, or indirectly through impacts to qubit coherence, negatively impacted by the inter-chip coupling technology. 

\begin{figure}
    \centering
    \includegraphics[width=\columnwidth]{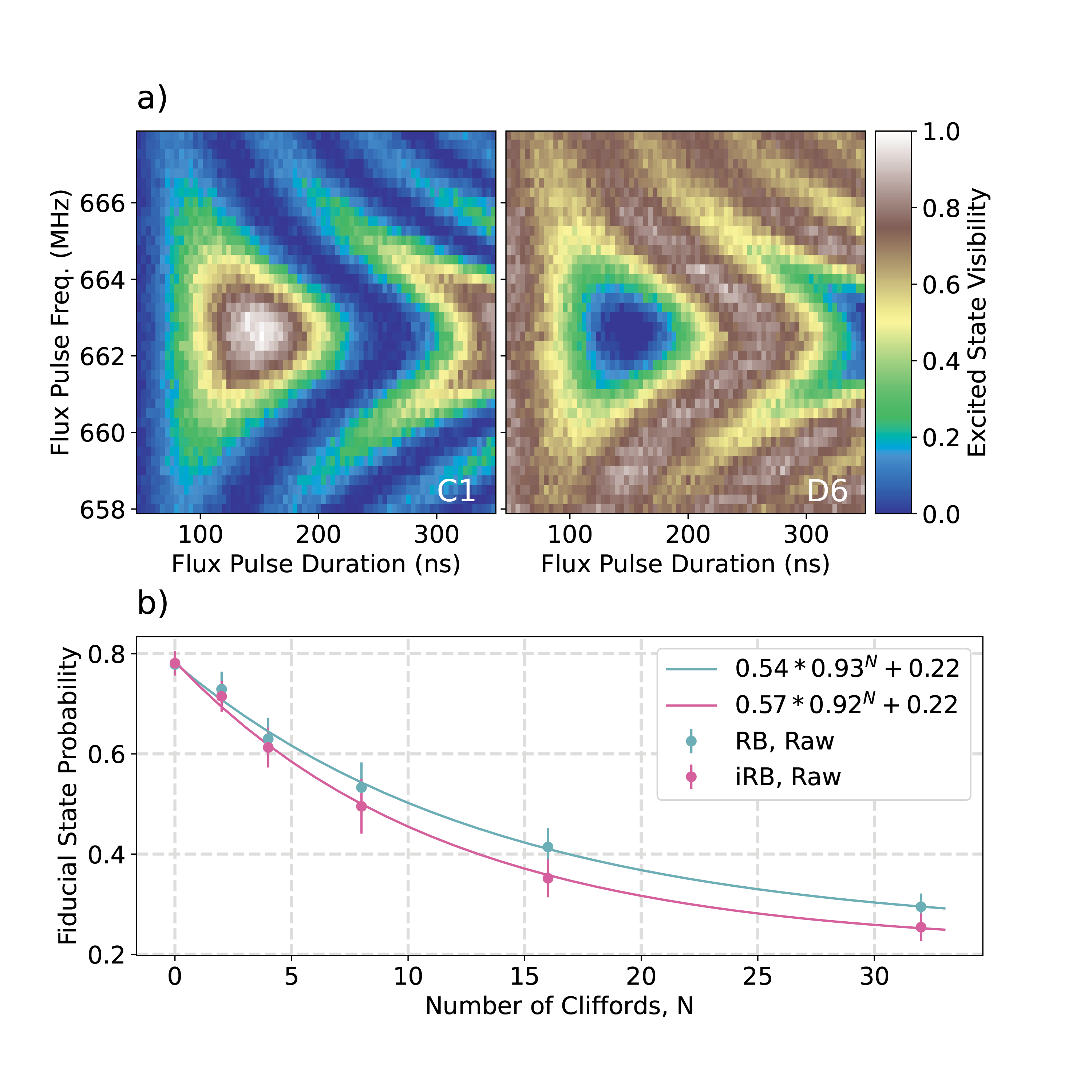}
    \caption{\label{fig:TwoQGateSummary} Two-qubit gate data for an iSWAP gate on the C1-D6 pair. (a) excited state visibility as a function of AC flux modulation frequency and pulse duration. Full population transfer occurs at a gate modulation frequency of 662.5 MHz and 152 ns. (b) RB and iRB benchmarking data. Fitting the exponential decay for the standard and interleaved benchmark gives an RB fidelity of 95.8\%$\pm$0.2\% and an iRB fidelity of 99.1\%$\pm$0.5\%.}
\end{figure}

Apart from four gates with two level systems which could explain the increased dephasing under modulation, CZ gate fidelities were above 90\% with 5 out of 12 gates demonstrating greater than 95\% measured fidelities. While Table~\ref{tab:Coherence Limited Fidelity} lists only CZ fidelities as CZ gates were within the AC flux control band for all pairs and they allowed a straightforward comparison to the coherence limited fidelity, the maximum gate fidelity measured was a 99.1\% $\pm$ 0.5\% iRB for an iSWAP gate on the C1-D6 pair, which we expand upon in Fig.~\ref{fig:TwoQGateSummary}. 

\begin{table*}
\caption{\label{tab:Coherence Limited Fidelity}%
Measured fidelity for CZ gates compared with coherence limited fidelity. The tunable qubit coherence times for the fixed - tunable pair with the tunable qubit under modulation, $\widetilde{\mathrm{T}}_1$ and $\widetilde{\mathrm{T}}_2$, are also listed.}
\begin{ruledtabular}
\begin{tabular}{lcccccc}
\textrm{Pair}&
\textrm{$\widetilde{\mathrm{T}}_1$ ($\mu$s)}&
\textrm{$\widetilde{\mathrm{T}}_2$  ($\mu$s)}&
\textrm{Coherence Limited Fidelity ($\%$)}&
\textrm{Measured Fidelity ($\%$)}&
\textrm{Gate Time (ns)}\\
\colrule
A0-B7 & 24.57$\pm$1.60 & 8.98$\pm$0.92 & 98.58$\pm$0.11 & 98.34$\pm$0.31 & 152\\
A1-B6 & 9.35$\pm$0.79 & 1.59$\pm$0.12 & 92.07$\pm$0.55 & 90.09$\pm$0.51 & 148\\
A3-B4 & 11.56$\pm$1.76 & 1.07$\pm$0.11 & 88.84$\pm$1.07 & 82.70$\pm$0.78 & 164\\
B0-C7 & 26.67$\pm$4.61 & 2.74$\pm$0.31 & 96.74$\pm$0.33 & 96.04$\pm$0.72 & 128\\
B2-C5 & 4.59$\pm$0.78 & 1.66$\pm$0.21 & 88.17$\pm$1.30 & 84.63$\pm$0.92 & 176\\
B3-C4 & 7.16$\pm$0.92 & 1.36$\pm$0.11 & 97.40$\pm$0.15 & 97.47$\pm$0.94 & 116\\
C0-D7 & 1.52$\pm$0.42 & 2.75$\pm$0.37 & 90.81$\pm$0.98 & 87.08$\pm$0.59 & 284\\
C1-D6 & 14.51$\pm$0.67 & 2.52$\pm$0.24 & 96.92$\pm$0.24 & 97.26$\pm$0.29 & 108\\
C2-D5 & 7.49$\pm$0.67 & 1.93$\pm$0.14 & 77.25$\pm$1.42 & 80.68$\pm$0.98 & 468\\
C3-D4 & 30.72$\pm$2.50 & 5.09$\pm$0.65 & 98.20$\pm$0.19 & 96.78$\pm$1.73 & 116\\
\end{tabular}
\end{ruledtabular}
\end{table*}
\subsection{\label{sec:bell} Multi-die Bell Inequality Violation}
We now turn our attention to assessing the viability of a future modular quantum processor based on these techniques. Importantly for this analysis, the inter-chip connections on our test device are established by unique qubits, and, in addition, qubits fabricated on the same chip are not coupled. We thus investigate the simultaneous quality of two-qubit connections, for the three disjoint pairs. This step is important for assessing functional challenges towards leveraging non-local quantum states in larger scale algorithms. Following a tradition established for multi-node or modular experimental efforts in superconducting qubits \cite{Narla2016,dickel2018chip,Campagne2018,zhong2019}, we design a test for the deterministic violation of a Bell inequality with this inter-chip platform. We describe a figure of merit $ \langle \mathcal{W}_{\Sigma} \rangle =\sum_k \langle W_k \rangle$, where $W_k$ is a witness to entanglement of connection $k$, applying the standard Bell observable for two-qubits,
\begin{equation}
    \mathcal{W} = Q S  +  RS +  RT  -  QT,
\end{equation}
with $Q=Z_n$, $R=X_n$, $S=\frac{X_m-Z_m}{\sqrt{2}}$, and $T=\frac{X_m+Z_m}{\sqrt{2}}$, taking qubits $\{n,m\}$ across an inter-chip connection. For $N$ disjoint Bell pairs, the total figure of merit is bounded by $\langle\mathcal{W}_{\Sigma}\rangle\leq 2N\sqrt{2}$ and signal above $\langle\mathcal{W}_{\Sigma}\rangle>2N$ certifies that the network supports genuine entanglement over at least one connection simultaneously. Moreover, investigating the individual Bell signals $\langle W_{k}\rangle$ can test entanglement over each connection independently.

\begin{figure}
    \includegraphics[width=\columnwidth]{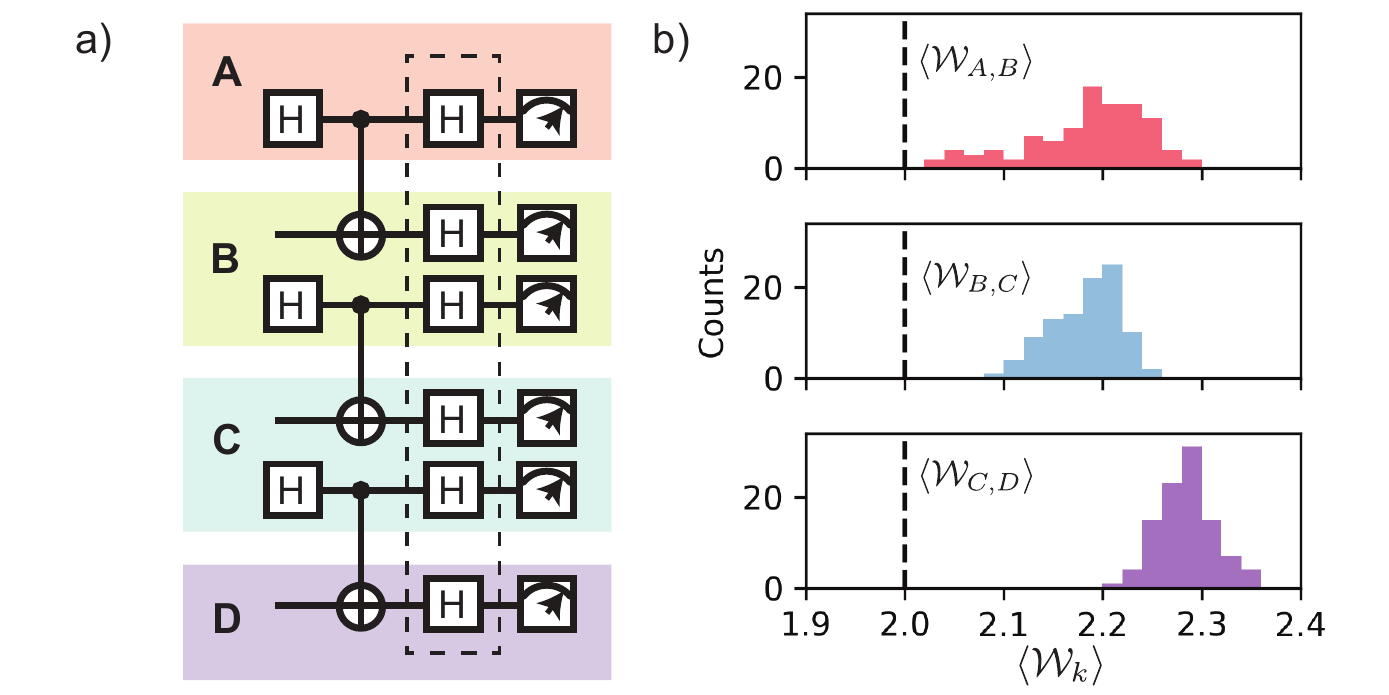}
    \caption{Simultaneous Bell Inequality Violations. (a) Disjoint, pairwise entanglement is generated across chip boundaries via CNOT operations on inter-chip couplers compiled to CZ gates, with the optional basis change for evaluating either $\langle ZZ \rangle$ (no final Hadamard) or $\langle XX \rangle$ (with final Hadamard). No error mitigation or readout correction schemes were applied. (b) Histograms of the average values $\langle W_k \rangle$ across 100 individual experimental runs using 10$^{4}$ shots in each measurement basis. The dashed line indicates the classical limit of 2.0. Because all three outcomes fall in the quantum limit, we conclude that the test platform supports simultaneous disjoint, pair-wise entanglement involving all four chips.}
    \label{fig:belltest}
\end{figure}

Our experimental procedure is shown in Fig.~\ref{fig:belltest}. We choose three connections that bridge all four chips in a disjoint pattern (A0-B7, B0-C7, C1-D6). We prepare the three pairs in an equal superposition of two-qubits: $|\Psi\rangle_k=(|00\rangle_k+|11\rangle_k)/\sqrt{2}$. Then, we measure the qubits in the $\langle ZZ \rangle$ basis or $\langle XX \rangle$ basis. A total of 100 experiments were run, having $10^{4}$ shots per basis, collecting measurement data simultaneously for all pairs. A summary of results is in Table~\ref{tab:bells}, where all three connections violate the Bell test by at least three standard deviations. With high confidence, therefore,  the test platform supports simultaneous disjoint, pair-wise entanglement involving all four chips. Additionally, our total figure of merit, $\langle \mathcal{W}_{\Sigma}\rangle=6.651\pm0.067$ exceeds the classical bound by nearly ten standard deviations.

\begin{table}
\caption{\label{tab:bells}Results for joint Bell test marginals and total observable.}
\begin{ruledtabular}
\begin{tabular}{ccc}
Observable & Qubits & Result \\
$\langle \mathcal{W}_{A,B} \rangle$ & A0-B7 & 2.184 $\pm$ 0.060 \\
$\langle \mathcal{W}_{B,C} \rangle$ & B0-C7 & 2.183 $\pm$ 0.034 \\
$\langle \mathcal{W}_{C,D} \rangle$ & C1-D6 & 2.284 $\pm$ 0.029 \\
\hline
$\langle \mathcal{W}_{\Sigma} \rangle$ & combined & 6.651 $\pm$ 0.067
\end{tabular}
\end{ruledtabular}
\end{table}

\section{\label{sec:Discussion}Discussion}

Concluding, we demonstrate that the flip-chip bonded, multi-die fabrication process with inter-chip coupler technology is capable of achieving high fidelity entanglement, including gate fidelities regularly exceeding 95\% and up to 99\% in the best case, and simultaneous entanglement between silicon dies violating the Bell test by over three standard deviations.

Future work should explore the potential benefits of this modular approach beyond the intrinsic advantages in regards to flexible device construction and yield.  This includes increased isolation between qubits on separate physical die, an important factor particularly in developing robust hardware suitable for near-term error correction schemes. Recently, the impacts of cosmic and background radiation on solid state devices have been of significant interest due to the correlated errors that result and the challenge these pose to fault tolerant quantum computing \cite{wilen2020correlated, vepsalainen2020impact, cardani2020reducing}. In this case and more generally, the physics of quasi-particle trapping and phonon propagation through superconducting qubit chips would be interesting to explore on multi-die devices. Phonons should collect on the boundaries of the individual die and not propagate to qubits on other dies, reducing correlated errors.

Finally, we note that the true impact of this technology will be in its integration with state-of-the-art processing architectures. With additional intra-chip circuitry and changes to the qubit topology on the individual QuICs, this technology can be extended to form a seamless modular quantum processor that is flexible in regards to the number and type of modules integrated and, with sensitivity to fabrication yield and intra-die cross-talk limited only by the module size, highly scalable. By enabling the fabrication of devices consisting of hundreds to thousands of qubits which are sufficiently isolated to mitigate correlated errors, this technology provides a clear path forward towards fault tolerant computing.

\section{\label{sec:Methods}Methods}
\subsection{\label{app:Fab} Fabrication and Bonding Process}
The carrier chip is composed of cavities etched in Si, coated with patterned superconducting metal, and indium bumps which after bonding form the superconducting connection between carrier and QuICs. Carrier chips are fabricated from high resistance Si wafers. The fabrication flow starts with a photolithography process to create cavity patterns on wafers followed by a Bosch etch (DRIE) step to fabricate 24 $\mu$m deep pockets with vertical sidewalls. The surface is then conformally coated with a 560 nm-thick Nb/MoRe stack, deposited by sputtering (PVD), to form a continuous superconducting shield. Vertical cavity sidewalls are confirmed to have a continuous metal film connecting the top surface to the cavity bottom surface. A thin layer of MoRe alloy is deposited on top of Nb film to seal the Nb surface, enabling an oxide-free metal-to-metal interface for reliable electrical connection between the Nb device layer and the In bumps in the bonding areas. 

The metal film stack is then patterned by a two-plane photolithography process followed by a reactive ion etching (RIE) step with a certain etching selectivity to the Si substrate. During the first exposure, focus and dose settings are selected to target the top wafer surface patterning, while in the second exposure settings are changed to target the cavity bottom surface only, which is 24 $\mu$m deeper. Once the Nb/MoRe stack etching is completed, a negative-tone photoresist lithography is used to transfer the In bump patterns onto the top metal surface by lift-off processing \cite{o2017superconducting}. Electron-beam evaporation is used to deposit a 6.5 $\mu$m thick indium layer, producing a high quality, easy to lift-off film. An automatic lift-off tool that uses a combination of chemical cleaning and physical energy produced by high-pressure jets removes the In film from non-patterned areas, completing the process. No Josephson junction fabrication steps are required as no active components are located on the carrier chip. Future designs could include transmons in the carrier chip as part of, for example, a quantum bus to provide longer-range bus coupling between qubits instead of the direct coupling employed in this initial design. 

To establish a superconducting connection between the carrier chip and the four separate QuIC chips, a flip chip bonder is used. Each QuIC device is precisely aligned and thermo-compression bonded to the carrier chip. Prior to bonding, both carrier and QuIC chip surfaces are solvent cleaned followed by an atmospheric downstream plasma cleaning to chemically clear surfaces of native oxides. This process also temporarily passivates the In bumps from oxidation and helps to generate a strong chemical bond between In bumps and the corresponding pads. The multi-chip bonding process consists of sequential bonding of four separate QuICs to the designated locations on the carrier chip. For each bonding, the carrier and QuIC chips are aligned to each other with a horizontal accuracy of better than $\pm$2.5 $\mu$m. After the horizontal alignment is completed, a vertical parallelism adjustment is done using auto-collimation and laser-levelling methods with an accuracy of $\pm$0.5 $\mu$m. The ensuing thermo-compression process consists of three different phases. In the first phase, force and temperature values are gradually increased and stabilized. In the second phase, the actual thermo-compression bonding takes place for two minutes. To prevent thermal aging of QuICs that are already bonded, the carrier chip temperature is maintained at 30$^\circ$C, while the QuICs are heated to 70$^\circ$C only during bonding. In the final phase, the stack is cooled to 30$^\circ$C with a nitrogen flow. The same process is repeated sequentially for all the QuICs. 

\subsection{\label{app:DevHam}Device Hamiltonian and Parametric Gates}

To design the device Hamiltonian, the circuit parameters were extracted using quasi-static electromagnetic simulations and a positive second order representation  \cite{scheer2018computational} was used to solve the linearized circuit. The non-linear effects of the Josephson junctions are subsequently accounted for through a perturbative treatment. The designed Hamiltonian parameters for this device are provided in table \ref{tab:devHam}, including the maximum and minimum $f_{01}$ transition frequencies over the flux bias tuning range, the anharmonicity at the maximum of the tuning range, $\eta=f_{12,\mathrm{max}}-f_{01,\mathrm{max}}$, the frequency of the readout resonator coupled to the qubit, $f_{\mathrm{RO}}$, and the qubit-readout resonator dispersive shift, $\chi_{\mathrm{q, RO}}$. The coupling rate between qubit pairs was designed to be $12$ MHz at a 3$\mu$m indium bump height for all couplers. 
\begin{table*}
\caption{\label{tab:devHam}Hamiltonian properties as designed. The qubit numbering corresponds to that shown in Fig.~\ref{fig:multiDieSchem}b, where there are only 4 unique design targets which are repeated in inverted order on the opposite side of the chip.}
\begin{ruledtabular}
\begin{tabular}{cccccccc}
Qubit & Flux Tunable? & $f_{01, \mathrm{max}}$ (MHz)&$f_{01, \mathrm{min}}$ (MHz)& $\eta$ (MHz)&$f_{\mathrm{RO}}$ (MHz)&$\chi_{\mathrm{q, RO}}/2\pi$ (MHz)\\ \hline
I & Fixed & 3654 & 3654 & -190 & 7232 & 0.80\\
II & Tunable & 5066 & 4266 & -200 & 7476 & 0.81\\
III & Fixed & 3714 & 3714 & -190 & 7273 & 0.85\\
IV & Tunable & 4946 & 4146 & -200 & 7425 & 0.81
\end{tabular}
\end{ruledtabular}
\end{table*}

Due to fabrication process variation, the Josephson junction width, and hence the Josephson energy, of each fabricated junction will differ slightly from the design target. Using the relationship between the room temperature conductance of a junction and its Josephson energy at cryogenic temperatures \cite{PhysRevLett.10.486}, a more accurate prediction for the device Hamiltonian can be obtained for fabricated devices using the same modelling process as during the initial device design, but replacing the target EJ values with the predicted EJ values from room temperature conductance measurements. Changes to the Josephson energy of the single junction in a fixed transmon or the two junctions in a DC-SQUID tunable transmon primarily impact the qubit frequencies, with little impact on qubit anharmonicities, readout resonator properties, or coupling rates. 

\begin{figure}
\centering
\includegraphics[width=\columnwidth]{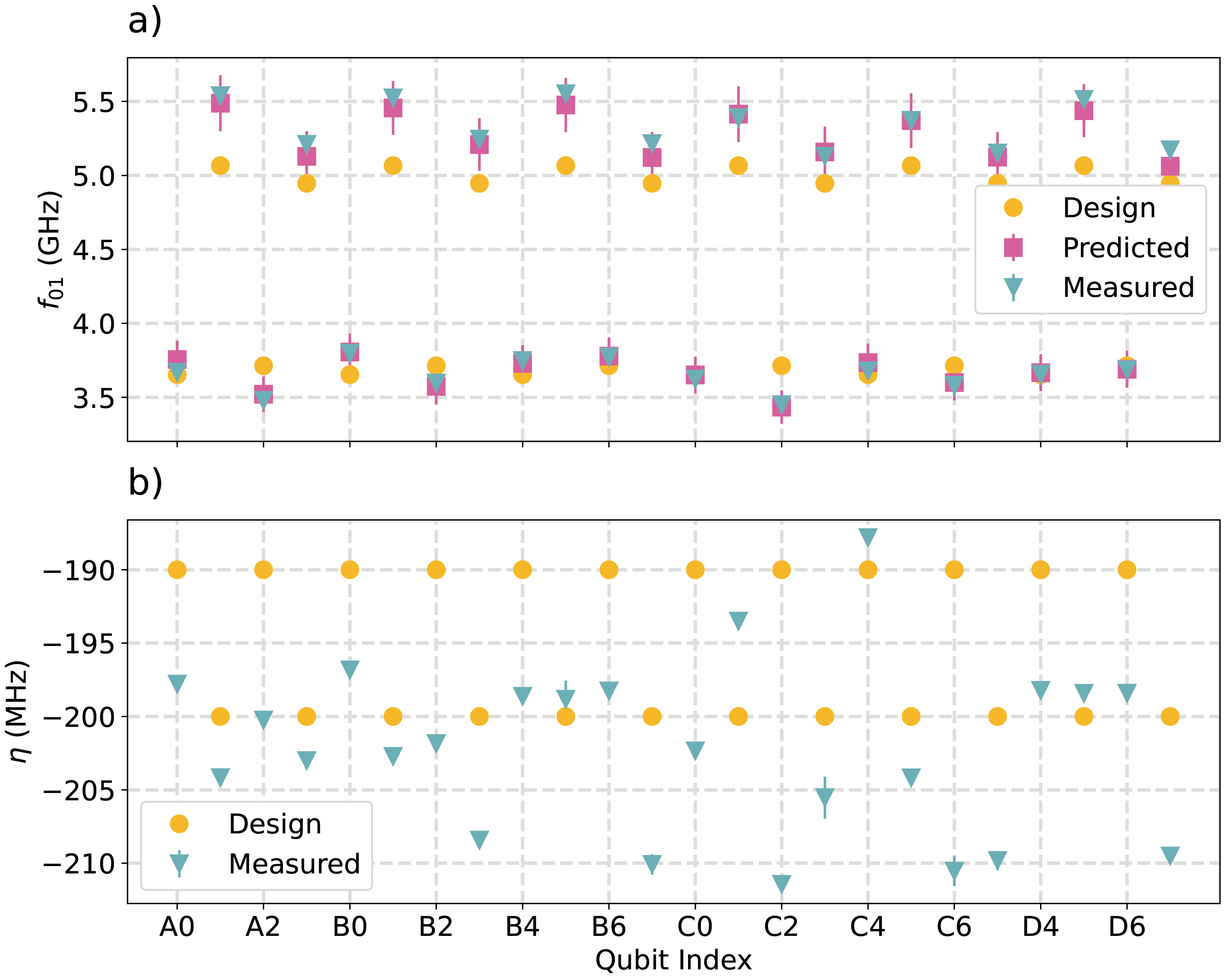}
\caption{Comparison of predicted and measured transition frequency, $f_{01}$ and anharmonicity, $\eta$ at zero applied flux bias. Note that while the design target is constant from one die to the next (all dies are designed to be identical), small variations in the predicted and measured qubit frequencies are present due to fabrication process variation, which can be measured at room temperature. Good agreement between the predicted and measured qubit frequencies, to within $\pm 108$ MHz, is obtained. Anharmonicities show a systematic offset from the design targets but are still within 11\% of the target value.}
\label{fig:HamSumm}
\end{figure}

We plot in Fig.~\ref{fig:HamSumm} the design target, predicted and measured $f_{01, \mathrm{max}}$, demonstrating agreement between the predicted frequencies and those measured cold, to within $\pm$ 108 MHz or 2.2\% in the worst case. The discrepancies are within the prediction error we expect due to uncertainty in the empirically determined linear coefficient relating room temperature conductance to inductance at cryogenic temperatures, and uncertainty in the conductance measurement itself. The qubit anharmonicities are compared with the design target and are accurate to within 11\%, demonstrating a systematic offset that will be corrected in future designs.

The device Hamiltonian is designed to enable parametric gates between one tunable ($T$) qubit coupled to one fixed qubit ($F$). In this scheme, an AC flux bias at RF frequency $f_p$ is applied to the tunable qubit around its parking flux bias. Under flux modulation, the transmon frequency oscillates at harmonics of the modulation frequency around its time-averaged frequency $\bar{f}_{T,01}$. Transmon frequency modulation gives rise to sidebands at frequencies $\bar{f}_{T,01}+kf_p$, separated by the modulation frequency around the average frequency. When the modulation frequency is tuned such as to align one sideband with the transition frequency of the fixed qubit, a coherent exchange takes place between the two qubits at a rate equal to the bare coupling strength renormalized by the sideband weight. When the tunable qubit is parked at the maximum of the tuning band, only even sidebands have a non-zero weight and the sideband $k=\pm2$ is used. Entangling gates are then enacted by modulating the tunable qubit at half the average detuning between the qubits' transition frequencies. To obtain the iSWAP gate, the interaction between states $|01\rangle$ and $|10\rangle$ is activated at the modulation frequency $f_p=|\bar{f}_{T,01}-f_{F,01}|/2\equiv\Delta/2$ (with the convention $|FT\rangle$). For the CZ gate, the interaction between $|11\rangle$ and $|02\rangle$ is activated at $f_p=(\Delta+\eta_T)/2$ (CZ$_{02}$) or between $|11\rangle$ and $|20\rangle$ at $f_p=(\Delta-\eta_F)/2$ (CZ$_{20}$). The gate time is adjusted to provide a $\pi$ rotation for iSWAP and $2\pi$ rotation for CZ between the corresponding two-qubit states.

\subsection{\label{app:g_analysis} Analysis of Bump Heights and Coupling Rates}

\begin{figure}
\centering
\setlength{\fboxsep}{0pt}%
\setlength{\fboxrule}{1pt}%
\fbox{\includegraphics[width=0.9\columnwidth]{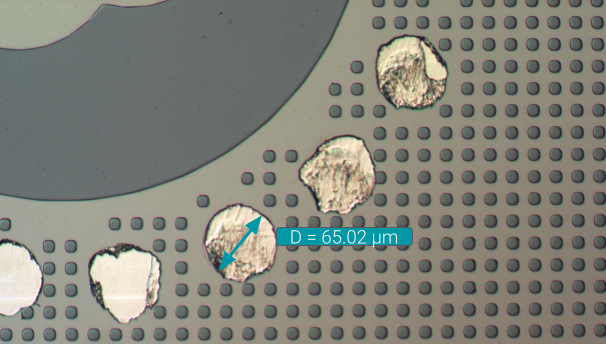}}
\caption{Magnified image of a region of the carrier chip showing the indium bumps post-bonding and post-shearing, a destructive process whereby a cut is made through the device separating the QuIC chips from the carrier chip. The diameter and height of the indium bumps are known pre-bonding, such that by measuring the bump diameter post-bonding, a rough estimate for the bump height can be obtained assuming the bump is cylindrical in both cases: $h_{\mathrm{post}} = h_{\mathrm{pre}} \left(D_{\mathrm{pre}}/D_{\mathrm{post}}\right)^2$.}
\label{fig:shearTestPics}
\end{figure}

After cryogenic tests were complete, the indium bump heights were measured at various locations across the device. This was done in a destructive manner by shearing the chip to separate the QuIC and the carrier chips, then measuring the indium bump diameter, as shown in Fig.~\ref{fig:shearTestPics}. Working under the assumption that before and after bonding the indium bumps are approximately cylindrical in shape, and with the diameter and height known before bonding, the diameter as measured after shearing provides an estimate for the bump height post-bonding. The measured coupling rates could then be compared with post-hoc simulated coupling rates computed for each coupler based on the measured bump heights. We obtain qualitative agreement between the simulated and measured coupling rates, however the measured coupling rates are approximately 20\% lower than expected from design. Future studies will look to address this discrepancy. Some of the reasons for the latter are the overestimation of the bump height size and inaccuracies in material properties assumed in the simulation. In addition the measured coupling values are effectively reduced by a term representing the next-nearest neighbor couplings to resonators, which exist in a higher band than the transmons.
\begin{table}
\caption{\label{tab:CouplingStrengths}%
Bare coupling rate, $g$, as measured from the qubit-qubit dispersive shift, $\chi_{qq}$ compared to simulated values given measured bump heights, $h$.
}
\begin{ruledtabular}
\begin{tabular}{lccc}
\textrm{Pair}&
\textrm{$h$ ($\mu$m)}&
\textrm{Meas. $g/2\pi$ (MHz)}&
\textrm{Sim. $g/2\pi$ (MHz)}\\
\colrule
A0-B7 & 2.18$\pm$0.13 & 13.26$\pm$0.59 & 16.86$\pm$1.05\\
A1-B6 & 2.03$\pm$0.08 & 14.89$\pm$1.00 & 18.44$\pm$0.79\\
A2-B5 & 1.96$\pm$0.18 & 15.72$\pm$0.96 & 19.32$\pm$1.96\\
A3-B4 & 1.94$\pm$0.29 & 16.46$\pm$1.07 & 19.60$\pm$3.15\\
B0-C7 & 1.95$\pm$0.27 & 14.34$\pm$0.44 & 19.42$\pm$2.89\\
B1-C6 & 1.89$\pm$0.18 & 14.91$\pm$0.62 & 20.11$\pm$2.12\\
B2-C5 & 1.69$\pm$0.14 & 15.40$\pm$0.54 & 22.62$\pm$2.07\\
B3-C4 & 1.76$\pm$0.08 & 16.63$\pm$1.81 & 21.29$\pm$1.04\\
C0-D7 & 1.77$\pm$0.04 & 14.54$\pm$0.87 & 21.11$\pm$0.51\\
C1-D6 & 1.69$\pm$0.01 & 18.94$\pm$0.39 & 22.29$\pm$0.07\\
C2-D5 & 1.55$\pm$0.01 & 18.88$\pm$1.70 & 24.52$\pm$0.16\\
C3-D4 & 1.60$\pm$0.11 & 18.92$\pm$0.43 & 23.70$\pm$1.66\\
\end{tabular}
\end{ruledtabular}
\end{table}

\subsection{\label{app:fid_summ} Impact of Inter-chip Couplers on Qubit Coherence and 2Q Gate Stability}

An important question to address is whether inter-chip coupling exposes qubits to additional loss or dephasing channels relative to standard intra-chip lateral couplers. As the electric field between the paddles of the coupler passes through vacuum rather than a silicon substrate, no additional dielectric losses are expected. Furthermore, the galvanic connection across the carrier chip is small enough relative to the 3-8 GHz band of interest that it can be treated as a lumped element and is not expected to produce any additional resonant coupling between qubits and the electromagnetic environment (chip modes, package modes, etc.) up to frequencies in excess of 15 GHz. 
\begin{figure}
\centering
\includegraphics[width=\columnwidth]{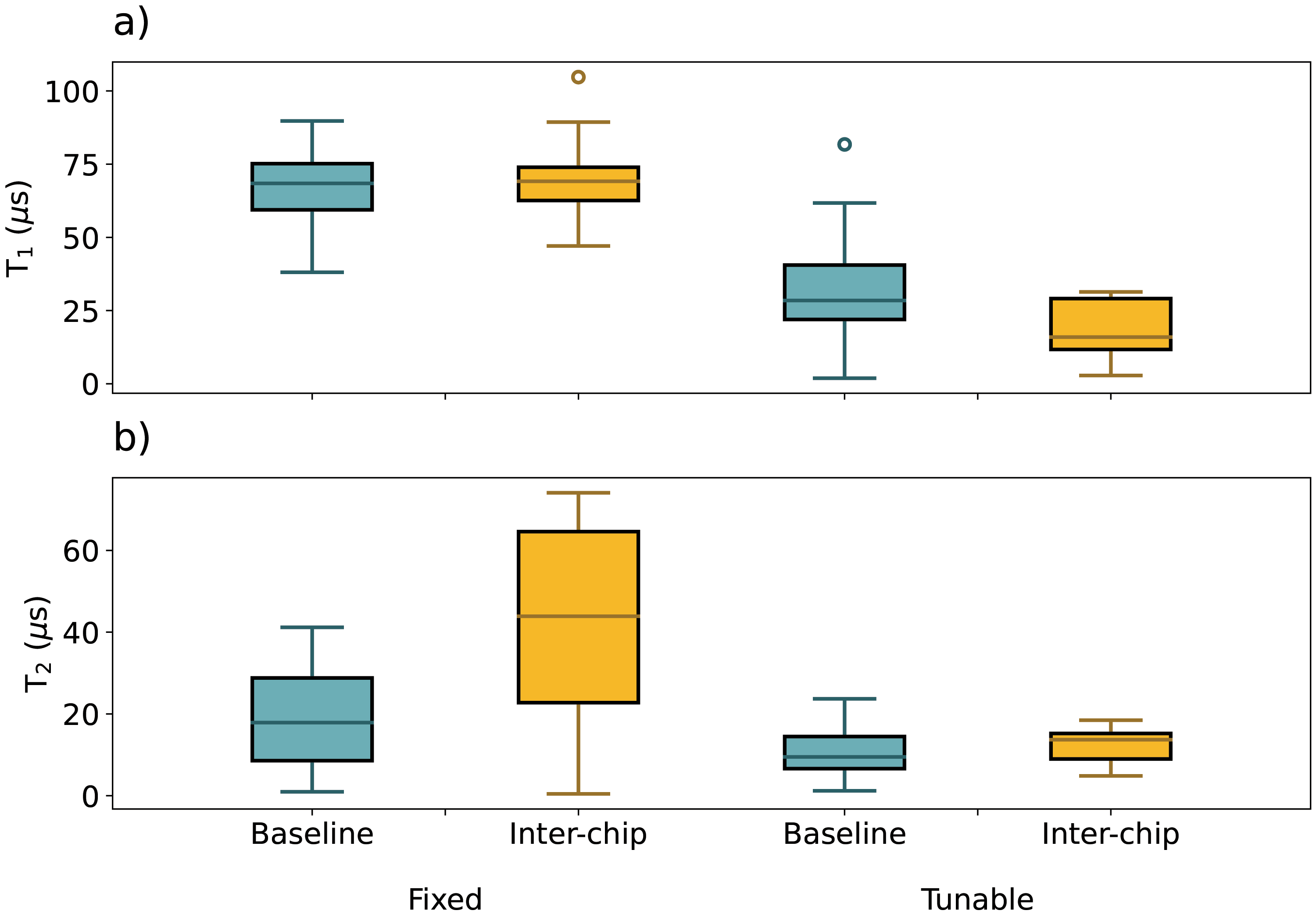}
\caption{Comparison of $\mathrm{T}_1$ and $\mathrm{T}_2$ for qubits with inter-chip coupling (inter-chip) compared to qubits on devices from similar wafers without inter-chip couplers (baseline). The baseline data includes 4 fixed qubits and 27 tunable qubits while the inter-chip data includes 12 fixed and 12 tunable qubits, and for each qubit the measured values are averaged over several measurements taken over the course of a week. The boxes plotted depict the four inter-quartile ranges for the distribution of average values with outliers shown as hollow circles. Tunable qubits have lower $\mathrm{T}_1$ and $\mathrm{T}_2$ due to the coupling of the flux bias line to the qubit and are thus plotted separately here.}
\label{fig:T1T2comp}
\end{figure}

It is thus not expected that these couplers should impact the qubit relaxation ($\mathrm{T}_1$) or dephasing ($\mathrm{T}_2$) times. This was reflected in experimental results, as shown in Fig.~\ref{fig:T1T2comp} where we compare $\mathrm{T}_1$ and $\mathrm{T}_2$ for a device with inter-chip coupling compared to devices from similar wafers with no coupling at all. No statistically significant difference in the $\mathrm{T}_1$ and $\mathrm{T}_2$ times was observed relative to the baseline.

\begin{figure}
\centering
\includegraphics[width=\columnwidth]{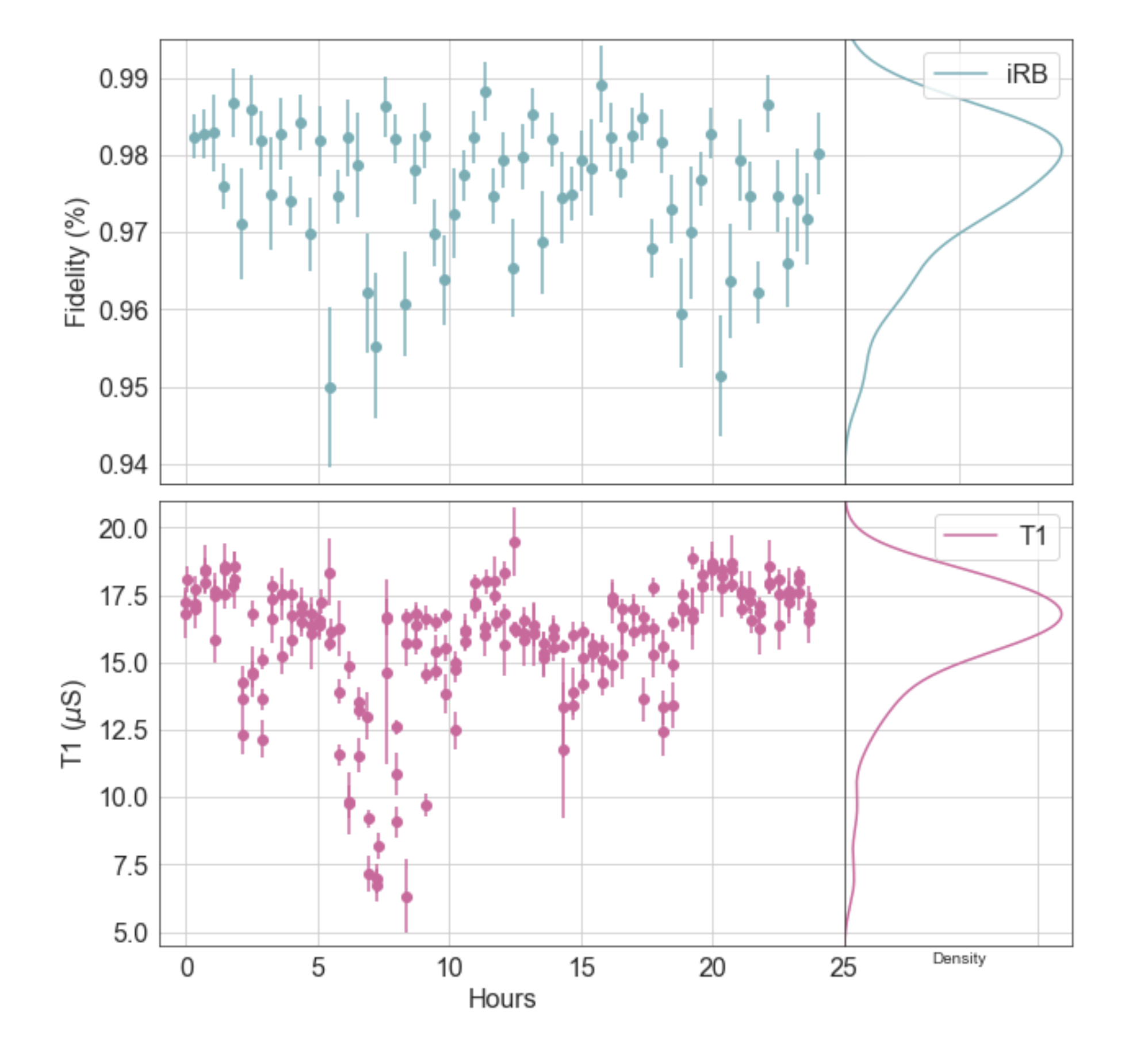}
\caption{Timeseries of the iRB fidelity and $\mathrm{T}_1$ of the tunable qubit (the lower of the tunable and fixed $\mathrm{T}_1$ and thus the limiting factor in regards to incoherent error) on the C1-D6 pair. The repeated measurement sequence calibrates readout parameters, parks the tunable qubit at its maximum frequency, measures the coherence time and benchmarks the gate. }
\label{fig:iRBtimeseries}
\end{figure}

In addition to comparing the coherence between inter-chip qubits and intra-chip qubits, we also assess the entangling gate stability and relaxation time, $\mathrm{T}_1$ over a period of 24 hours, as plotted for the C1-D6 pair in Fig.~\ref{fig:iRBtimeseries}. For each data point plotted, the qubits are re-tuned, readout is calibrated, and the two-qubit gate is benchmarked using interleaved randomized benchmarking. Qubit re-tuning is done by parking the tunable qubit at its maximum frequency and calibrating the gate pulses for the fixed and tunable qubit. Readout calibration involves preparing ground and excited states to update the classifier that will be used to discriminate single shot measurements. $\mathrm{T}_1$ decay is monitored through repeated coherence measurements taken after re-tuning and readout calibrations and before benchmarking. Temporal fluctuations show a drop in the gate fidelity when the $\mathrm{T}_1$ falls below 10 $\mu$s, but generally it remains stable to within four percentage points with a distribution centered around 98\%. Fluctuations in $\mathrm{T}_1$ is an active research topic in the field of superconducting qubits \cite{vepsalainen2020impact, cardani2020reducing}. 

\begin{acknowledgments}
The authors would like to thank A. Grassellino, A. Romanenko, L. Cardani, and R. McDermott for insightful discussions on the effects of cosmic and background radiation on qubit coherence. We thank the Rigetti fabrication team for manufacturing the device, the Rigetti technical operations team for fridge build out and maintenance, the Rigetti cryogenic hardware team for providing the chip packaging, the Rigetti quantum engineering team for guidance during measurement and data analysis, and the Rigetti control systems and embedded software teams for creating the Rigetti AWG control system.

This material is based upon work supported by Rigetti Computing and the Defense Advanced Research Projects Agency (DARPA) under agreement No. HR00112090058 and DARPA under IAA 8839, Annex 114.
\end{acknowledgments}

\bibliography{references}% Produces the bibliography via BibTeX.

\begin{thebibliography}{10}

\bibitem{PRXQuantum.2.017002}
Awschalom, D. et~al.
\newblock Development of Quantum Interconnects (QuICs) for Next-Generation
  Information Technologies.
\newblock \emph{PRX Quantum} \textbf{2}, 017002 (2021).

\bibitem{steffen2013deterministic}
Steffen, L. et~al.
\newblock Deterministic quantum teleportation with feed-forward in a solid
  state system.
\newblock \emph{Nature} \textbf{500}, 319--322 (2013).

\bibitem{chou2018deterministic}
Chou, K.~S. et~al.
\newblock Deterministic teleportation of a quantum gate between two logical
  qubits.
\newblock \emph{Nature} \textbf{561}, 368--373 (2018).

\bibitem{wan2019quantum}
Wan, Y. et~al.
\newblock Quantum gate teleportation between separated qubits in a trapped-ion
  processor.
\newblock \emph{Science} \textbf{364}, 875--878 (2019).

\bibitem{hensen2015loophole}
Hensen, B. et~al.
\newblock Loophole-free Bell inequality violation using electron spins
  separated by 1.3 kilometres.
\newblock \emph{Nature} \textbf{526}, 682--686 (2015).

\bibitem{olmschenk2009quantum}
Olmschenk, S. et~al.
\newblock Quantum teleportation between distant matter qubits.
\newblock \emph{Science} \textbf{323}, 486--489 (2009).

\bibitem{moehring2007entanglement}
Moehring, D.~L. et~al.
\newblock Entanglement of single-atom quantum bits at a distance.
\newblock \emph{Nature} \textbf{449}, 68--71 (2007).

\bibitem{humphreys2018deterministic}
Humphreys, P.~C. et~al.
\newblock Deterministic delivery of remote entanglement on a quantum network.
\newblock \emph{Nature} \textbf{558}, 268--273 (2018).

\bibitem{monroe2014large}
Monroe, C. et~al.
\newblock Large-scale modular quantum-computer architecture with atomic memory
  and photonic interconnects.
\newblock \emph{Physical Review A} \textbf{89}, 022317 (2014).

\bibitem{ritter2012elementary}
Ritter, S. et~al.
\newblock An elementary quantum network of single atoms in optical cavities.
\newblock \emph{Nature} \textbf{484}, 195--200 (2012).

\bibitem{Zhong_2020}
Zhong, C. et~al.
\newblock Proposal for Heralded Generation and Detection of Entangled
  Microwave–Optical-Photon Pairs.
\newblock \emph{Physical Review Letters} \textbf{124} (2020).

\bibitem{krastanov2021}
Krastanov, S. et~al.
\newblock Optically-Heralded Entanglement of Superconducting Systems in Quantum
  Networks (2021).

\bibitem{PhysRevLett.120.200501}
Campagne-Ibarcq, P. et~al.
\newblock Deterministic Remote Entanglement of Superconducting Circuits through
  Microwave Two-Photon Transitions.
\newblock \emph{Phys. Rev. Lett.} \textbf{120}, 200501 (2018).

\bibitem{PhysRevLett.125.260502}
Magnard, P. et~al.
\newblock Microwave Quantum Link between Superconducting Circuits Housed in
  Spatially Separated Cryogenic Systems.
\newblock \emph{Phys. Rev. Lett.} \textbf{125}, 260502 (2020).

\bibitem{axline2018demand}
Axline, C.~J. et~al.
\newblock On-demand quantum state transfer and entanglement between remote
  microwave cavity memories.
\newblock \emph{Nature Physics} \textbf{14}, 705--710 (2018).

\bibitem{leung2019deterministic}
Leung, N. et~al.
\newblock Deterministic bidirectional communication and remote entanglement
  generation between superconducting qubits.
\newblock \emph{npj Quantum Information} \textbf{5}, 1--5 (2019).

\bibitem{zhong2020deterministic}
Zhong, Y. et~al.
\newblock Deterministic multi-qubit entanglement in a quantum network.
\newblock \emph{arXiv preprint arXiv:2011.13108}  (2020).

\bibitem{kurpiers2018deterministic}
Kurpiers, P. et~al.
\newblock Deterministic quantum state transfer and remote entanglement using
  microwave photons.
\newblock \emph{Nature} \textbf{558}, 264--267 (2018).

\bibitem{Chen_2020}
Chen, M.-C. et~al.
\newblock Demonstration of Adiabatic Variational Quantum Computing with a
  Superconducting Quantum Coprocessor.
\newblock \emph{Physical Review Letters} \textbf{125} (2020).

\bibitem{Foxen_2020}
Foxen, B. et~al.
\newblock Demonstrating a Continuous Set of Two-qubit Gates for Near-term
  Quantum Algorithms.
\newblock \emph{Physical Review Letters} \textbf{125} (2020).

\bibitem{negirneac2020highfidelity}
Negîrneac, V. et~al.
\newblock High-fidelity controlled-Z gate with maximal intermediate leakage
  operating at the speed limit in a superconducting quantum processor (2020).

\bibitem{sung2020realization}
Sung, Y. et~al.
\newblock Realization of high-fidelity CZ and ZZ-free iSWAP gates with a
  tunable coupler (2020).

\bibitem{stehlik2021tunable}
Stehlik, J. et~al.
\newblock Tunable Coupling Architecture for Fixed-frequency Transmons (2021).

\bibitem{wilen2020correlated}
Wilen, C.~D. et~al.
\newblock Correlated Charge Noise and Relaxation Errors in Superconducting
  Qubits (2020).

\bibitem{vepsalainen2020impact}
Veps{\"a}l{\"a}inen, A.~P. et~al.
\newblock Impact of ionizing radiation on superconducting qubit coherence.
\newblock \emph{Nature} \textbf{584}, 551--556 (2020).

\bibitem{cardani2020reducing}
Cardani, L. et~al.
\newblock Reducing the impact of radioactivity on quantum circuits in a
  deep-underground facility.
\newblock \emph{arXiv preprint arXiv:2005.02286}  (2020).

\bibitem{8614500}
{Brink}, M., {Chow}, J.~M., {Hertzberg}, J., {Magesan}, E. \& {Rosenblatt}, S.
\newblock Device challenges for near term superconducting quantum processors:
  frequency collisions.
\newblock In \emph{2018 IEEE International Electron Devices Meeting (IEDM)},
  pp. 6.1.1--6.1.3 (2018).

\bibitem{dickel2018chip}
Dickel, C. et~al.
\newblock Chip-to-chip entanglement of transmon qubits using engineered
  measurement fields.
\newblock \emph{Physical Review B} \textbf{97}, 064508 (2018).

\bibitem{rosenberg20173d}
Rosenberg, D. et~al.
\newblock 3D integrated superconducting qubits.
\newblock \emph{npj quantum information} \textbf{3}, 1--5 (2017).

\bibitem{foxen2017qubit}
Foxen, B. et~al.
\newblock Qubit compatible superconducting interconnects.
\newblock \emph{Quantum Science and Technology} \textbf{3}, 014005 (2017).

\bibitem{brecht2016multilayer}
Brecht, T. et~al.
\newblock Multilayer microwave integrated quantum circuits for scalable quantum
  computing.
\newblock \emph{npj Quantum Information} \textbf{2}, 1--4 (2016).

\bibitem{reagor2018demonstration}
Reagor, M. et~al.
\newblock Demonstration of universal parametric entangling gates on a
  multi-qubit lattice.
\newblock \emph{Science advances} \textbf{4}, eaao3603 (2018).

\bibitem{bertet2006parametric}
Bertet, P., Harmans, C. \& Mooij, J.
\newblock Parametric coupling for superconducting qubits.
\newblock \emph{Physical Review B} \textbf{73}, 064512 (2006).

\bibitem{niskanen2007quantum}
Niskanen, A. et~al.
\newblock Quantum coherent tunable coupling of superconducting qubits.
\newblock \emph{Science} \textbf{316}, 723--726 (2007).

\bibitem{didier2018parametric}
Didier, N., Sete, E.~A., da~Silva, M.~P. \& Rigetti, C.
\newblock Analytical modeling of parametrically modulated transmon qubits.
\newblock \emph{Phys. Rev. A} \textbf{97}, 022330 (2018).

\bibitem{hong2020demonstration}
Hong, S.~S. et~al.
\newblock Demonstration of a parametrically activated entangling gate protected
  from flux noise.
\newblock \emph{Physical Review A} \textbf{101}, 012302 (2020).

\bibitem{nersisyan2019manufacturing}
{Nersisyan}, A. et~al.
\newblock Manufacturing low dissipation superconducting quantum processors.
\newblock In \emph{2019 IEEE International Electron Devices Meeting (IEDM)},
  pp. 31.1.1--31.1.4 (2019).

\bibitem{PhysRevLett.125.240503}
Xu, Y. et~al.
\newblock High-Fidelity, High-Scalability Two-Qubit Gate Scheme for
  Superconducting Qubits.
\newblock \emph{Phys. Rev. Lett.} \textbf{125}, 240503 (2020).

\bibitem{PhysRevApplied.10.054062}
Yan, F. et~al.
\newblock Tunable Coupling Scheme for Implementing High-Fidelity Two-Qubit
  Gates.
\newblock \emph{Phys. Rev. Applied} \textbf{10}, 054062 (2018).

\bibitem{Magesan_2012}
Magesan, E. et~al.
\newblock Efficient Measurement of Quantum Gate Error by Interleaved Randomized
  Benchmarking.
\newblock \emph{Physical Review Letters} \textbf{109} (2012).

\bibitem{knill2008randomized}
Knill, E. et~al.
\newblock Randomized benchmarking of quantum gates.
\newblock \emph{Physical Review A} \textbf{77}, 012307 (2008).

\bibitem{Narla2016}
Narla, A. et~al.
\newblock Robust Concurrent Remote Entanglement Between Two Superconducting
  Qubits.
\newblock \emph{Phys. Rev. X} \textbf{6}, 031036 (2016).

\bibitem{Campagne2018}
Campagne-Ibarcq, P. et~al.
\newblock Deterministic Remote Entanglement of Superconducting Circuits through
  Microwave Two-Photon Transitions.
\newblock \emph{Phys. Rev. Lett.} \textbf{120}, 200501 (2018).

\bibitem{zhong2019}
Zhong, Y.~P. et~al.
\newblock Violating {Bell}’s inequality with remotely connected
  superconducting qubits.
\newblock \emph{Nature Physics} \textbf{15}, 741--744 (2019).

\bibitem{o2017superconducting}
O'Brien, W. et~al.
\newblock Superconducting caps for quantum integrated circuits.
\newblock \emph{arXiv preprint arXiv:1708.02219}  (2017).

\bibitem{scheer2018computational}
Scheer, M.~G. \& Block, M.~B.
\newblock Computational modeling of decay and hybridization in superconducting
  circuits.
\newblock \emph{arXiv preprint arXiv:1810.11510}  (2018).

\bibitem{PhysRevLett.10.486}
Ambegaokar, V. \& Baratoff, A.
\newblock Tunneling Between Superconductors.
\newblock \emph{Phys. Rev. Lett.} \textbf{10}, 486--489 (1963).

\end{thebibliography}

\end{document}